\documentstyle[preprint,prb,aps]{revtex}
\begin{document}
\draft
\title{Giant persistent current in free-electron model with flat Fermi
surface.}
\author{E. V. Tsiper, A. L. Efros}
\address{Department of Physics, University of Utah, Salt Lake City, Utah,
84112}
\date{\today}
\maketitle

\begin{abstract}

For the first time the persistent current in a 2D free-electron system
has been calculated analytically.  The tight binding model is
considered on a square lattice with filling factor 1/2.  The array has
a shape of rectangle with boundary conditions in both directions
twisted by $2\pi\phi_x$ and $2\pi\phi_y$.  The components of the twist
are associated with two components of the magnetic flux in torus
geometry.  An analytical expression is obtained for the energy and for
the components of the persistent current (PC) at a given flux and
temperature.  It is shown that at zero temperature the PC density is
proportional to the vector potential with the coefficient which does
not depend on the size of the system.  This happens because the Fermi
surface for a square lattice at filling factor 1/2 is flat.  Both the
energy and the PC are periodic functions of the two flux components
with the periods $\phi_0/q$ and $\phi_0/s$ where $\phi_0=hc/e$, and
$q$ and $s$ are integers which depend on the aspect ratio of the
rectangle.  The magnitude of PC is the same as in superconductors.
Therefore, a 3D system constructed from a macroscopic number of
isolated coaxial cylinders at zero temperature reminds the London's
superconductor.  It exhibits the quantization of trapped flux as well
as the Meissner effect.  However, all the phenomena are of a
mesoscopic nature.  The critical field $H_c$ decays with an effective
size of the system, $H_c\sim 1/R_{ef}$.  The magnitude of PC decays
with $T$ as $\exp(-\pi TR_{ef}/2at)$, where $t$ is the hopping
amplitude and $a$ is the lattice constant.

\end{abstract}

\section{Introduction}

Persistent current (PC) in mesoscopic structures\cite{kohn,buttiker}
has been extensively studied during the last decade both
experimentally\cite{levy,chandra,mailly} and theoretically.  The
theoretical investigations concentrated on a role of different degrees
of disorder\cite{alt,atl} and on the role of the interaction between
electrons\cite{berk,efr}.

The PC is a reaction of a system to an applied flux $\Phi$, or,
equivalently, it can be described as a change of the energy of the
system due to twisted boundary conditions. In a two-dimensional system
which forms a cylinder the twisted conditions mean that the wave
function of a system acquires a factor $\exp(i2\pi\Phi/\phi_0)$ with a
circulation of one electron around the axis of the cylinder.  Here
$\phi_0=hc/e$ is the flux quanta.

The flux $\Phi$ is related to the tangential component $A$ of the vector
potential $\Phi=2\pi RA$, where $R$ is the radius of the cylinder.
For a system with Galilean invariance the following simple statement
is correct.  The energy of a state with a given value of tangential
component $P$ of the total momentum depends on $A$ as

\begin{equation}
E(P,A)=E_0-\frac{1}{2M}\left(P-\frac{Ne}{c}A\right)^2,
\end{equation}
where $N$ is the number of electrons and $M=Nm$ is their total mass,
$m$ being the mass of one electron.  The 2D current density $j$ for a
state with fixed $P$ is

\begin{equation}
j_P=-{c\over S}\left({\partial E\over\partial A}\right)_P
  =-{ne^2\over mc}A+{eP\over mS},
\label{london}
\end{equation}
where $S$ is the area of the cylinder surface and $n=N/S$.  At $P=0$
Eq.~(\ref{london}) reminds the London equation for a superconducting
current.  In this case $n$ should be the superfluid density.

A general derivation of Eq.~(\ref{london}), given above, is misleading
because PC should be defined as a current in the ground state rather
than in a state with fixed $P$.  In two- or three-dimensional systems
of free electrons the derivative of the energy with respect to $A$
cannot be taken in such a simple way because the intervals of $\Phi$,
where branches of spectrum with different $P$ change each other in the
ground state, tend to zero with increasing system size.

For electrons in a periodic potential the situation is typically
similar.  The derivative of energy with respect to the flux is large
for a given branch.  However, different branches replace each other in
the ground state at such small intervals of $\Phi$ that the derivative
taken at a given total quasimomentum $P$ does not reflect properties
of the ground state.  Scalapino et al.\cite{scal} considered a tight
binding model on a two-dimensional square lattice.  Their computations
show that at filling factor $\nu=1/4$ the first level crossing occurs
at $\Phi\sim 1/L$, where $L$ is the size of the system.  Their general
conclusion is that the superfluid density, as found from the relation
between $j$ and $A$, is zero for free electrons in the tight binding
model.  We show in this paper that this is not always the case.

Namely, we consider a 2D system of free electrons on a square lattice
in a tight binding approximation at filling factor $\nu=1/2$.  The
shape of the system is assumed to be a rectangle with arbitrary aspect
ratio.  We demonstrate below that at $T=0$ the two-dimensional PC
density does not depend on the size of the system and has a form:

\begin{equation}
j=-\frac{4}{\pi^2}\ \frac{ne^2}{mc}(A-A_0).
\label{result}
\end{equation}
Here $m=\hbar^2/2ta^2$ is the electron mass, $t$ being the
nearest-neighbor hopping energy. The two-dimensional density is
determined as $n=1/2a^2$, where $a$ is the lattice constant.  For
simplicity, we consider a system of spinless fermions.  The
generalization to the case of non-interacting fermions with spin is
straightforward.

The constant $A_0$ shows that the minimum of energy occurs not at zero
flux.  In contrast to Eq.~(\ref{london}), Eq.~(\ref{result}) describes
PC in the ground state of the system which is a periodic function of
$\Phi$ with period $\phi_0/q$.  Eq.~(\ref{result}) is valid within the
interval $0<\Phi<\phi_0/q$, or $0<A<\phi_0/2\pi Rq$, and is to be
repeated periodically for other values of flux.  Here $q$ is an
integer which depends on the aspect ratio of the cylinder and on the
type of the boundary conditions imposed in the direction of the
cylinder axis.  The first term is shown to be independent on the
aspect ratio.

We found PC to have an order of magnitude of the London current.  Note
that this result gives substantially larger PC than it is prescribed
in the ballistic regime by the so-called $M$-channel approximation
(see Ref.\onlinecite{gefen1} and references therein).  Namely, our
exact solution gives PC larger by a factor of $\sqrt{L}$ for the
$L\times L$ square.  This discrepancy is due to the fact that in the
case of flat Fermi surface all transverse channels are coherent.

Considering the 3D system constructed of a large number of coaxial
closely packed 2D cylinders we show that it mimics the Meissner effect
and the quantization of flux trapped in the opening.

These properties appear since the Fermi surface at $\nu=1/2$ is flat
and no branch crossings occur in large intervals of $\Phi$.  Say, for
a square array no branch crossing occurs in the whole interval
$0<\Phi<\phi_0$, which means that $q=1$.

In fact, we are discussing a mesoscopic effect.  The expression
Eq.~(\ref{result}) is valid only at mesoscopically small temperature,
and the ideal diamagnetism occurs for mesoscopically small values of
magnetic field:

\begin{eqnarray}
T&<& T_c\sim \frac{a}{R_{ef}}t,
\label{Tc}\\
H&<& H_c\sim\frac{a}{R_{ef}}\sqrt{\frac{t}{a^2b}},
\label{Hc}
\end{eqnarray}
where $b$ is the spacing between neighboring coaxial cylinders.  The
effective size $R_{ef}$ is given by

\begin{equation}
R_{ef}=sD=q2\pi R=\sqrt{sq2\pi RD},
\label{Ref}
\end{equation}
where $D$ is the length of the cylinders and $s$ and $q$ are integers
determined by the aspect ratio $2\pi R/D$ (see below).  Thus, this
system can be classified as a ``mesoscopic superconductor.''

Note that the average distance between energy levels in a 2D system is
proportional to $1/R^2$.  The $1/R$ behavior in the above equations 
is also a result of the flat Fermi surface at $\nu=1/2$.  As it is
seen from the calculations, all relevant interlevel distances are of
the order of $(a/R)t$, rather than $(a/R)^2t$.

Since both $T_c$ and $H_c$ vanish at large $R$, there are no real
critical phenomena in this model system.

\section{Calculation of PC at zero temperature}

Consider a rectangle of $L_x\times L_y$ lattice sites with periodic
boundary conditions twisted in both directions by $2\pi\Phi_x/\phi_0$
and $2\pi\Phi_y/\phi_0$.  This corresponds to a toroidal geometry
where $\Phi_x$ is the flux through the crossection of the torus and
$\Phi_y$ is the flux through the opening.

The single-electron energies have the form

\begin{equation}
\epsilon(n_x,n_y,\phi_x,\phi_y)=-2J\left\{
  \cos\left[\frac{2\pi}{L_x}(n_x-\phi_x)\right]+
  \cos\left[\frac{2\pi}{L_y}(n_y-\phi_y)\right]
\right\},
\label{epsilon1}
\end{equation}
where we introduce dimensionless $\phi_{x,y}=\Phi_{x,y}/\phi_0$ to
simplify the notation.  The values of integer quantum numbers $n_x$
and $n_y$ are restricted to the rectangle $|n_{x,y}|\leq L_{x,y}/2$
(first Brillouin zone).  To find the energy of the ground state one
has to sum $\epsilon(n_x,n_y)$ over the values $\{(n_x,n_y)\}$ inside
the Fermi surface.

In principle, the calculation of PC can be performed either at
constant number of particles $N$ or at constant value of chemical
potential $\mu$.  Generally speaking, these two definitions are not
equivalent.  It is important to note that such a problem does not exist
at $\nu=1/2$ at even numbers of $L_x,L_y$ at least.  As one can see
from the Eq.~(\ref{epsilon1}), every single-electron energy changes sign
under the transformation $n_x,n_y\rightarrow n_x+L_x/2,n_y+L_y/2$. It
follows, that at $\nu=1/2$ due to the electron-hole symmetry the
chemical potential $\mu$ is zero at any value of flux and at any
temperature. Thus, if the flux changes at $\mu=0$, the number
of particles in the ground state of the system does not change and if
the flux changes at a given number of particles such that $\nu=1/2$,
the chemical potential does not change.

Let us define the Fermi ``surface'' (FS) in two-dimensional $n_x,n_y$
space by the equation

\begin{equation}
\epsilon(n_x,n_y,\phi_x,\phi_y)=0,
\label{mu}
\end{equation}
considering $n_x,n_y$ as continuous variables.  It is easy to see
that the FS forms a rhomb at any value of flux.  Change in the flux
produces a shift of the FS as a whole without changing its shape.

First, let us consider for simplicity a square sample, $L_x=L_y$.  The
FS forms a square shown in Fig. 1(a).  At $\phi_{x,y}=0$ some of the
allowed single-electron states lie exactly at the sides of this
square.  All the states inside the square and 1/2 of the states at the
sides of the square are occupied.  All the states at the sides have
the same energy so the occupation numbers of these states are not
defined, while the many-electron ground state is degenerate.

The degeneracy is lifted at infinitazimally small values of $\phi$.
Suppose that $\phi_y=0$ and $\phi_x>0$.  Then FS is shifted to the
right (See Fig.\ 1(a)).  All occupation numbers become defined.
Namely, the states at the right side of initial square get occupied
and those at the left side become empty.  Note that the occupation
numbers of as many as $2L$ states change when $\phi_x$ crosses zero.

It is easy to see that the occupation numbers are constant throughout
the interval $0<\phi_x<1$.  The total energy decreases with
$\phi_x$ and then increases again.  At $\phi_x=1$ all electrons
jump one step to the right and the Fermi surface restores its original
position with respect to the lattice of integer numbers $(n_x,n_y)$.
The total energy thus returns to the same value as at $\phi_x=0$.

It follows that the total quasimomentum of electron system in the
ground state does not change through all this interval and no branch
crossing occurs.  Then the sum over the occupied states can be easily
evaluated:

\begin{eqnarray}
E(\phi_x,\phi_y)&=&\sum_{n_y=1}^{L/2-1}\sum_{n_x=-L/2+n_y+1}^{L/2-n_y}
    \left[\epsilon(n_x,n_y)+\epsilon(n_x,-n_y)\right]+
  \sum_{n_x=-L/2+1}^{L/2}\epsilon(n_x,0)\\
&=&8t\ \text{Re}\frac{e^{2\pi i/L}}{(e^{2\pi i/L}-1)^2}
  \left[\left(1+e^{2\pi i/L}\right)e^{-i2\pi\phi_x/L}+
     2e^{-i2\pi\phi_y/L}\right]
  \nonumber
\end{eqnarray}
This expression is exact in the region $0<\phi_x\pm\phi_y<1$.  In the
limit of large $L$, the $\phi$-dependent part of the energy,
$\delta E(\phi_x,\phi_y)=E(\phi_x,\phi_y)-E(0,0)$, can be written in
the form:

\begin{equation}
\delta E(\phi_x,\phi_y)=8t\left[\phi_y^2-\phi_x(1-\phi_x)\right].
\label{energy1}
\end{equation}
Repeating Eq.~(\ref{energy1}) periodically one gets the expression
valid in the whole plane $(\phi_x,\phi_y)$:

\begin{equation}
  \delta E(\phi_x,\phi_y)=4t\left[
     \left(\left\{\phi_+\right\}-\frac{1}{2}\right)^2+
     \left(\left\{\phi_-\right\}-\frac{1}{2}\right)^2
  -\frac{1}{2}\right],
\label{energy2}
\end{equation}
where $\phi_\pm=\phi_x\pm\phi_y$, and $\{...\}$ denotes the fractional
part of the number, defined as a difference between the number and the
largest integer less than it.

Fig.\ 2 shows the energy $\delta E(\phi_x,\phi_y)$ as given by
Eq.~(\ref{energy2}).  The positions of energy minima form a square
lattice shifted from the origin:

\begin{equation}
(\phi_x,\phi_y)=\left(\frac{1+i+j}{2},\frac{i-j}{2}\right),
\label{minima1}
\end{equation}
with arbitrary integer $i$ and $j$.

The point $\phi_x=\phi_y=0$ corresponds to a maximum of energy, in the
same way as in 1D case with even number of electrons.  At this point
the derivatives $d\delta E/d\phi_{x,y}$ are discontinuous.  Such
behavior appears as a result of lifting of the degeneracy of the
states at the Fermi surface.

The PC at $T=0$ can be found as the derivative of the total energy
with respect to flux:

\begin{eqnarray}
I_{x,y}&=&-c\frac{\partial E}{\partial\Phi_{x,y}}=
  -\frac{c}{\phi_0}\left(
    \frac{\partial E}{\partial\phi_+}\pm
    \frac{\partial E}{\partial\phi_-}\right)\nonumber\\
&=&-8\frac{ct}{\phi_0}\left[
    \left(\{\phi_+\}-\frac{1}{2}\right)\pm
    \left(\{\phi_-\}-\frac{1}{2}\right)\right]
\label{current2}
\end{eqnarray}

The magnitude of $\delta E(\phi_x,\phi_y)$ and
$I_{x,y}(\phi_x,\phi_y)$ as given by
Eqs.~(\ref{energy2}), (\ref{current2}) is independent on the size $L$
of the square.  Such a large magnitude results from the fact that in
the region with no branch crossings (or, with no electron changing its
state) all electrons together contribute to the current.

It may seem that the aspect ratio $L_x/L_y=1$ is crucial for the
effect.  In the next section we calculate PC at finite temperature for
arbitrary aspect ratio $L_x=sK$, $L_y=qK$ with mutually-prime integers
$s$ and $q$.  We assume macroscopic limit $K\rightarrow\infty$.  It is
useful to generalize $\phi_\pm$ for a rectangular sample as
\begin{equation}
\phi_\pm=q\phi_x\pm s\phi_y.
\label{phipm}
\end{equation}

In the limit $T=0$ we find
\begin{eqnarray}
I_x&=&-\frac{8}{sq}\frac{ct}{\phi_0/q}\left[
  \left(\left\{\phi_+\right\}-\frac{1}{2}\right)+
  \left(\left\{\phi_-\right\}-\frac{1}{2}\right)\right]\nonumber\\
\label{current3}\\
I_y&=&-\frac{8}{sq}\frac{ct}{\phi_0/s}\left[
  \left(\left\{\phi_+\right\}-\frac{1}{2}\right)-
  \left(\left\{\phi_-\right\}-\frac{1}{2}\right)\right]\nonumber
\end{eqnarray}

The flux-dependent part of the energy can be restored from
Eq.~(\ref{current3}):

\begin{equation}
  \delta E(\phi_x,\phi_y)=\frac{4t}{sq}\left[
     \left(\left\{\phi_+\right\}-\frac{1}{2}\right)^2+
     \left(\left\{\phi_-\right\}-\frac{1}{2}\right)^2
  -\frac{1}{2}\right],
\label{energy3}
\end{equation}
This result is a generalization of Eq.~(\ref{energy2}) to an arbitrary
aspect ratio $s/q$ of the rectangular sample.

As follows from Eqs.~(\ref{current3}), (\ref{energy3}), the energy and
current as functions of flux do depend on the aspect ratio.  However,
they do not depend on the system size, if the aspect ratio is kept
constant.

The result Eqs.~(\ref{current3}), (\ref{energy3}), can be understood
from Fig.\ 1(b), which is drawn for the case $L_x/L_y=2/3$.  Contrary
to the Fig.\ 1(a), there are now points $(n_x,n_y)$ closer to the Fermi
surface than one lattice spacing.  However, there is still a
regularity in their positions.  Namely, as the Fermi surface shifts
with flux, the points enter the Fermi sea in groups.  Consider, for
example, the same case as above: $\phi_y=0$ and $\phi_x>0$.  As seen
in Fig.\ 1(b), the branch crossings occur only at $\phi_x=2\pi/3$,
$4\pi/3$, and $2\pi$.  In terms of $\phi_\pm$ this corresponds to
integer $\phi_\pm=3\phi_x=1$, 2, and 3.  These values of flux are
determined by $s$ and $q$ and do not change with the size of the
system.  The number of points in each group, in tern, is proportional
to the size of the system, so the corresponding contribution to the
current is large.

At $\Phi_y=0$ Eq.~(\ref{current3}) gives

\begin{equation}
I_x=-\frac{16}{sq}\frac{ct}{\phi_0/q}
  \left(\left\{\frac{\Phi_x}{\phi_0/q}\right\}-\frac{1}{2}\right).
\label{current}
\end{equation}

Up to now we are discussing the torus geometry.  To come to a cylinder
geometry one has to formulate the boundary conditions in the direction
of the cylinder axis, chosen as $y$.  In what follows we assume
periodic boundary conditions in this direction with $\Phi_y=0$.  This
leads to Eq.~(\ref{current}) for a total current through the cylinder.
As another option me may impose the condition that the wave function
is zero at the edges of the cylinder.  It can be shown that in this
case the second term in Eq.~(\ref{current}) changes while the first
term remains intact.

Note that both energy and current are periodic functions of flux with
period $\phi_0/q$ rather than $\phi_0$.

Taking into account that the current density $j_x=I_x/(aL_y)$ and that
the vector potential $A_x=\Phi_x/(aL_x)$ one obtains
Eq.~(\ref{result}) with the first term independent of $s$ and $q$.

\section{PC at finite temperature}

We start with the equation
\begin{equation}
I_x=-\frac{c}{\phi_0}\sum_{n_x=0}^{sK-1}\sum_{n_y=0}^{qK-1}
  \frac{\partial\epsilon(n_x,n_y)}{\partial\phi_x}
  \frac{1}{1+\exp(\epsilon(n_x,n_y)/T)}.
\label{Ix}
\end{equation}
It is convenient to rewrite the single electron energy in the form

\begin{equation}
\epsilon(n_x,n_y)=-4t
  \cos\left(\frac{\pi}{sqK}(n_+-\phi_+)\right)
  \cos\left(\frac{\pi}{sqK}(n_--\phi_-)\right),
\label{epsilon2}
\end{equation}
where $n_\pm=qn_x\pm sn_y$ and $\phi_\pm$ are given by
Eq.~(\ref{phipm}).  Using $\partial/\partial\phi_x=
q(\partial/\partial\phi_+ +\partial/\partial\phi_-)$ we find that
the current has two terms,

\begin{equation}
I_x=\frac{1}{s}(I_++I_-),
\label{Ix1}
\end{equation}
where

\begin{equation}
I_\pm=-sq\frac{c}{\phi_0}\sum_{n_x=0}^{sK-1}\sum_{n_y=0}^{qK-1}
  \frac{\partial\epsilon(n_x,n_y)}{\partial\phi_\pm}
  \frac{1}{1+\exp(\epsilon(n_x,n_y)/T)}
\label{Ipm}
\end{equation}

The idea of our calculation is to transform Eq.~(\ref{Ipm}) in such a
way that the internal sum gives PC of 1D problem with effective
temperature and effective flux.  For this purpose we use the identity:

\begin{equation}
\sum_{n_x=0}^{sK-1}\sum_{n_y=0}^{qK-1}f(n_x,n_y)=
  \sum_{m=0}^{sK-1}\sum_{d=0}^{q-1}\sum_{k=0}^{K-1}f(m+d+sk,d+qk).
\label{identity}
\end{equation}
This identity is valid for any function $f(n_x,n_y)$ periodic in $n_x$
and $n_y$ with periods $sK$ and $qK$ respectively.  Then $I_+$ can be
written in the form

\begin{equation}
I_+=-\frac{4\pi ct}{K\phi_0}
  \sum_{m=0}^{sK-1}\sum_{d=0}^{q-1}\sum_{k=0}^{K-1}\frac{
    \sin\left(\frac{\pi}{sqK}(n_+-\phi_+)\right)
    \cos\left(\frac{\pi}{sqK}(n_--\phi_-)\right)
  }{
    1+\exp\left[-\frac{4t}{T}
      \cos\left(\frac{\pi}{sqK}(n_+-\phi_+)\right)
      \cos\left(\frac{\pi}{sqK}(n_--\phi_-)\right)\right]}.
\label{Ip}
\end{equation}
Similar expression can be written for $I_-$.  Note that
$n_-=qm+(q-s)d$ does not depend on $k$, while $n_+=(qm+qd+sd)+2sqk$
does depend on $k$.  Therefore, the current $I_+$ can be written as

\begin{equation}
I_+=\sum_{m=0}^{sK-1}\sum_{d=0}^{q-1}{\cal I}_+(m,d),
\label{Ip1}
\end{equation}
where ${\cal I}_+(m,d)$ denote the internal sum over $k$,

\begin{equation}
{\cal I}_+(m,d)=-\frac{4\pi ct}{K\phi_0}
  \frac{2T}{\widetilde T}\sum_{k=0}^{K-1}\frac{
    \sin\left(\frac{2\pi}{K}(k-\widetilde\phi_+)\right)
  }{
    1+\exp\left[-\frac{2t}{\widetilde T}
      \cos\left(\frac{2\pi}{K}(k-\widetilde\phi_+)\right)\right]}.
\label{calIp}
\end{equation}
The sum in Eq.~(\ref{calIp}) describes the PC in 1D system with
effective temperature and effective flux given by

\begin{eqnarray}
\widetilde T(m,d)&=&\frac{T}
  {2\cos\left(\frac{\pi}{sqK}(qm+(q-s)d-\phi_-)\right)}\nonumber\\
\label{effective}\\
\widetilde\phi_+(m,d)&=&\frac{\phi_+-qm-qd-sd}{2sq}\nonumber
\end{eqnarray}
Using the Poisson summation formula (see Ref.\ \onlinecite{gefen}),
one obtains

\begin{equation}
{\cal I}_+(m,d)=\frac{8\pi cT}{\phi_0}\sum_{l=1}^{\infty}
  \frac{\cos(l\pi K/2)}{\sinh(l\pi\widetilde TK/2t)}
  \sin(2\pi l\widetilde\phi_+)
\label{1Dres}
\end{equation}

Performing the summation over $m$ and $d$ in Eq.~(\ref{Ip1}) we note
that $\widetilde T$ is a smooth function of $m/K$ and $d/K$.  However,
$\sin(2\pi l\widetilde\phi_+)$ has an oscillatory behavior for some
$l$, so that the contribution of the corresponding harmonics vanishes
in the limit $K\rightarrow\infty$.  The oscillatory behavior is absent
if $l$ is an integer multiple of $2sq$.  For these $l$, the sum over
$m$ can be transformed into integral via $p=(\pi/sK)m$, while the sum
over $d$ simply gives a factor $q$.  Thus, one obtains

\begin{equation}
I_\pm=\sum_{l=1}^{\infty}A_l\sin(2\pi l\phi_\pm),
\label{Ipres}
\end{equation}
where

\begin{equation}
A_l=sqKT\frac{2c}{\phi_0}\int_0^\pi\frac{dp}
  {\sinh(l\pi sqKT/2t\sin p)}
\label{Al}
\end{equation}

For the PC in $x$-direction one has from Eq.~(\ref{Ix1})
\begin{equation}
I_x=\frac{1}{s}\sum_{l=1}^\infty A_l
  \left[\sin(2\pi l\phi_+)+\sin(2\pi l\phi_-)\right].
\label{Ixres}
\end{equation}

Similar calculation gives
\begin{equation}
I_y=\frac{1}{q}\sum_{l=1}^\infty A_l
  \left[\sin(2\pi l\phi_+)-\sin(2\pi l\phi_-)\right].
\label{Iyres}
\end{equation}

Eqs.~(\ref{Al}), (\ref{Ixres}) give the Fourier series expansion of
PC at any temperature.  Expansion of $A_l$ at small $KT/t$ yelds

\begin{equation}
A_l\approx\frac{8t}{\phi_0}\frac{1}{l\pi}
\label{Alzero}
\end{equation}
In this case Eqs.~(\ref{Ixres}), (\ref{Iyres}), and (\ref{Alzero})
give the Fourier series expansion of the zero-temperature result
Eq.~(\ref{current3}).

In the opposite limit, $KT/t\gg1$, the amplitudes of the harmonics
decay as

\begin{equation}
A_l\approx\frac{8c}{\phi_0\sqrt{l}}\sqrt{sqKTt}
  \exp\left(-\frac{l\pi sqKT}{2t}\right),
\label{Alinf}
\end{equation}
so that PC is dominated by its lowest harmonic.  When $\Phi_y=0$ one
has

\begin{equation}
I_x\approx\frac{1}{sq}\frac{16c}{\phi_0/q}
  \left(\frac{R_{ef}Tt}{a}\right)^{1/2}
  \exp\left(-\frac{\pi R_{ef}T}{2at}\right)
  \sin\left(\frac{\Phi_x}{\phi_0/q}\right).
\label{Ixinf}
\end{equation}

\section{Low temperature magnetic properties}

In this section we study magnetic properties of a quasi-3D system
constructed of a macroscopic number of closely packed coaxial
cylinders assuming that the temperature is very low.  Then the
connection between flux and current for each cylinder is given by
Eq.~(\ref{current}).  For the sake of simplicity we assume that the
cylinders are long, such that the circumference of the internal
cylinder $2\pi R=aL_x$ is much larger than $D=aL_y$.  The distance
between the internal and external cylinders is supposed to be much
less than $R$.  We assume further that all the cylinders have the same
ratio $L_x/L_y=s/q$.  One can imagine a small change either in $L_x$
and $L_y$ of adjacent cylinders or in their lattice constant.

The second term in Eq.~(\ref{current}) appears since zero flux does
not correspond to the minimum of energy.  It may lead to an appearance
of a spontaneous flux in this system.  This idea has been put forward
by Wohlleben et al, and Szopa and Zipper, Ref.\ \onlinecite{poland0},
and then studied in details in Ref.\ \onlinecite{poland}.  These
authors considered a cylinder constructed from isolated 1D rings.
Loss and Martin\cite{loss} argued that in a single 1D ring no symmetry
breaking can occur, but their arguments are restricted to 1D case.

In this paper we concentrate on the first term in
Eq.~(\ref{current}).  It is an analog of the London current in
superconductors and it creates a strong diamagnetism in a quasi-3D
system described above.  Suppose that an external magnetic field
$H_{ext}$ is applied to the system and that there is a solenoid
creating flux $\Phi_{ext}$ inside the internal cylinder.

Let $\Phi_k$ be the total flux inside cylinder $k$, where $k=1$ for
the internal cylinder and $k=N$ for the external one.  The flux obeys
the equation
\begin{equation}
\Phi_k-\Phi_{k-1}=
  2\pi Rb\left(H_{ext}+\frac{4\pi}{cD}\sum_{i=k}^NI(\Phi_i)\right).
\label{difference}
\end{equation}
Here $b$ is the distance between adjacent cylinders which we assume to
be of the order of the lattice constant $a$.  Since the thickness
$d=Nb$ is supposed to be much less than $R$ we have neglected that the
radii of cylinders are slightly different.  The right hand side of
Eq.~(\ref{difference}) describes the flux through the area between the
$k$-th and $(k-1)$-th cylinders created by the external field and outer
cylinders.

The following condition should be added to this finite difference
equation: 

\begin{equation}
\Phi_1-\Phi_{ext}=
  \pi R^2\left(H_{ext}+\frac{4\pi}{cD}\sum_{i=1}^NI(\Phi_i)\right).
\label{condition}
\end{equation}

If $\Phi_k$ is a smooth function of $k$ one can transform
Eq.~(\ref{difference}) into differential equation

\begin{equation}
\frac{d^2\Phi}{dr^2}=\frac{\phi_0/q}{\lambda^2}
  \left(\left\{\frac{\Phi}{\phi_0/q}\right\}-\frac{1}{2}\right).
\label{differential}
\end{equation}

Here $\lambda$ is the analog of the London penetration depth

\begin{equation}
\lambda^{-2}=\frac{4\pi}{b}\frac{16t}{\phi_0^2}
  =\frac{16e^2n_3}{\pi mc^2},
\label{lambda}
\end{equation}
where $n_3=1/2ba^2$ is the 3D electron density.

Eq.~(\ref{condition}) transforms into the boundary condition at $r=R$:
\begin{equation}
\Phi(R)-\Phi_{ext}=\frac{R}{2}\left.\frac{d\Phi}{dr}\right|_{r=R}.
\label{condition2}
\end{equation}

The second boundary condition reads
\begin{equation}
\left.\frac{d\Phi}{dr}\right|_{r=R+d}=2\pi RH_{ext}.
\label{condition3}
\end{equation}
One can use differential equation if $\lambda\gg b$.

Eq.~(\ref{differential}) can also be obtained by minimizing total
energy with respect to flux.  The total energy consists of two parts.
First is the energy of magnetic field in the space between cylinders.
The magnetic field can be expressed through $d\Phi/dr$ using
Eq.~(\ref{difference}) as

\begin{equation}
\frac{d\Phi}{dr}=2\pi RH(r).
\label{H}
\end{equation}

The second part is the internal energy of 2D electron gas.  This
energy per cylinder is given by Eq.~(\ref{energy3}) at $\phi_y=0$.
Thus, one gets for the total energy

\begin{equation}
E_{total}=\frac{1}{8\pi}\frac{D}{2\pi R}
  \int\left(\frac{d\Phi}{dr}\right)^2dr +
\int\delta E(\Phi)\frac{dr}{b}.
\label{Etotal}
\end{equation}

Minimizing this expression with respect to $\Phi(r)$ and taking into
account that $d\delta E/d\Phi=(-1/c)I(\Phi)$, where $I(\Phi)$ is given
by Eq.~(\ref{current}), one obtains Eq.~(\ref{differential}).

The Eq.~(\ref{differential}) is nonlinear since it contains the
fractional part $\{\Phi/(\phi_0/q)\}$ which makes the right-hand side
periodic.  However, it becomes linear if the total drop of the flux
inside the system is smaller than $\phi_0/q$.  If $H_{ext}=0$ the
solution of the linearized equation with boundary conditions
(\ref{condition2}), (\ref{condition3}) in the case $R\gg d\gg \lambda$
is

\begin{equation}
\Phi(r)=\Phi_n+(\Phi_{ext}-\Phi_n)\frac{2\lambda}{R}
  \exp\left(-\frac{r-R}{\lambda}\right).
\label{noH}
\end{equation}
Here

\begin{equation}
\Phi_n=\frac{\phi_0}{q}\left(n-\frac{1}{2}\right).
\label{Phi_n}
\end{equation}

One can see that the flux inside the cylinder with $d\gg\lambda$ may
take only quantized values $\Phi_n$ with arbitrary integer $n$.  Note
that there is no zero flux among the allowed values of the frozen flux
$\Phi_n$.  This is because zero flux does not correspond to a minimum
of the total energy at zero temperature. The solution Eq.~(\ref{noH})
is obtained in the linear approximation and it is valid if
$(\Phi_{ext}-\Phi_n) 2\lambda/R <\phi_0/q$. The physics of this result
is that the inner cylinders carry a current which creates a favorable
flux for the rest of the system.

If the system is in an external magnetic field $H_{ext}$, the solution
is
 
\begin{equation}
\Phi(r)=\Phi_n+2\pi R\lambda H_{ext}
  \exp\left(-\frac{R+d-r}{\lambda}\right).
\label{flH}
\end{equation}
or in terms of magnetic field defined by Eq.~(\ref{H})   
\begin{equation}
H(r)=H_{ext}\exp\left(-\frac{R+d-r}{\lambda}\right)
\label{Hext}
\end{equation}

In this case the cylinders near external surface carry current which
screens magnetic field inside the system and adjusts the total flux to
$\Phi_n$.  The solution is valid if $2\pi R\lambda H_{ext}<\phi_0/q$.
This condition is equivalent to Eq.~(\ref{Hc}). It has a simple
interpretation. The loss in the total energy due to the ideal Meissner
effect is of the order of $H_{ext}^2RDb$ per cylinder.  The gain
in the energy of a cylinder due to the adjusted flux is of the order
of $t/sq$ (see Eq.~(\ref{energy3})).  At large field the loss becomes
larger than the gain and the field penetrates into the system.  This
is the origin of a ``mesoscopic'' critical field.  Note that the
relation $H_c^2RDb\sim t/sq$ is also equivalent to
Eq.~(\ref{Hc}).

It follows from the results of the previous section that
zero-temperature approximation is good if $sqKT/t=T\sqrt{sq2\pi
RD}/at\ll1$.  This is the same condition as Eq.~(\ref{Tc}).  At larger
temperatures the penetration depth $\lambda$ increases as $\exp(\pi
TR_{ef}/4at)$ and eventually reaches the thickness $d$ of the cylinder,
gradually destroying strong diamagnetism.

\section{Conclusions}

Finally we have presented a model which mimics in a mesoscopic scale
some properties of superconductors, such as Meissner effect and
quantization of flux, though the physics of the model does not involve
any electron pairing.  The flux quanta in the model is $\phi_0/q$
where $q$ is determined by the aspect ratio of the system.

Since the range of temperature and magnetic field for these phenomena
shrinks to zero in a macroscopic system, one should not expect any
phase transitions.  However, for a mesoscopic system this range is not
necessarily small.  Let us assume a hypotetic 3D layered system with
very weak interaction between layers and flat two-dimensional Fermi
surface.  Then it follows from Eqs.~(\ref{Tc}), (\ref{Hc}) that the
temperature range is up to 12K and the range of $H_{ext}$ is up to 240
gauss for a system with $R_{ef}=3\times10^{-5}$cm,
$a=b=3\times10^{-8}$cm, and $t=1$eV.  In a system with disorder the
obvious condition for these phenomena is that the elastic mean free
path is smaller than the size $R_{ef}$.

Our model ignores electron-electron interaction.  We hope that it is
not important at large $t$.  Our modeling of small interacting systems
up to 18 electrons show the same value of the PC at $t$ immediately
above the Wigner crystal quantum melting point\cite{tsiper}.

\section{Acknowledgments}
The authors thank D.E. Khmelnitskii for helpful discussions.  The work
was supported by QUEST of UCSB, subcontract KK3017.

\begin{figure}
\caption{The Fermi surface at $\nu=1/2$.  The points represent allowed
integer values of $n_x$ and $n_y$ inside the first Brillouin zone.
The dashed lines show the Fermi surface at zero flux.  The solid lines
are the Fermi surface shifted by flux.  The aspect ratio $L_x/L_y=1$
(a) and $L_x/L_y=2/3$ (b).}
\end{figure}

\begin{figure}
\caption{Lines of constant $\delta E(\phi_x,\phi_y)$ as given by
Eq.~(\protect\ref{energy3}) for square sample $s=q=1$ (a), and for
rectangle with $s=2$ and $q=3$ (b).  Note the difference in
periodicities.}
\end{figure}

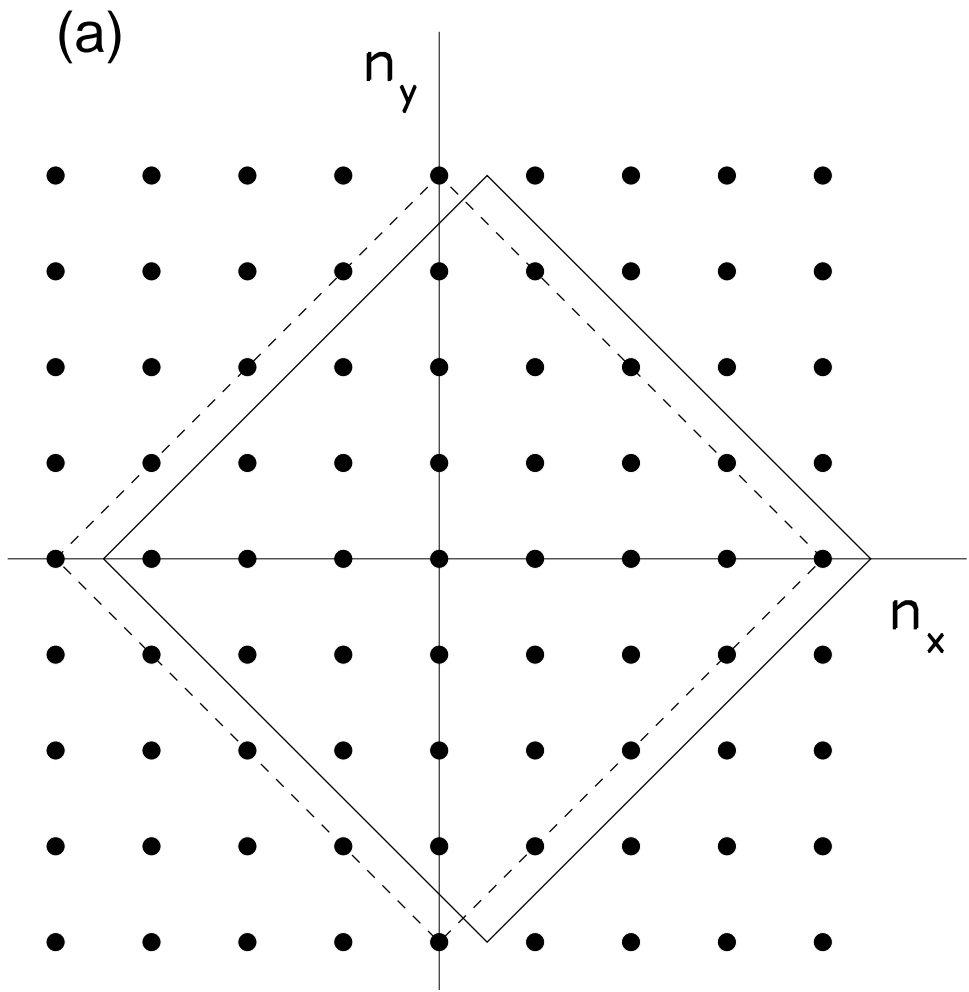
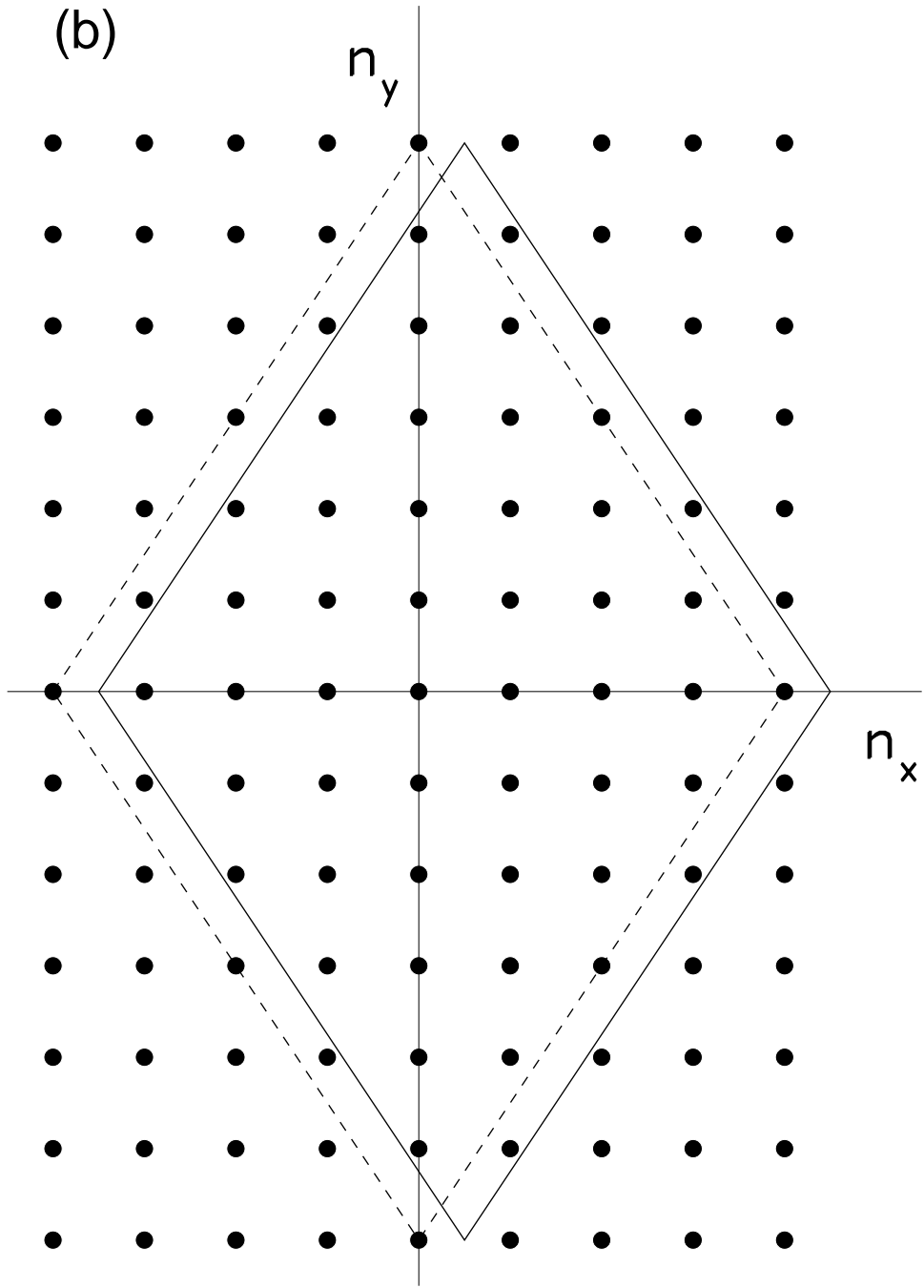
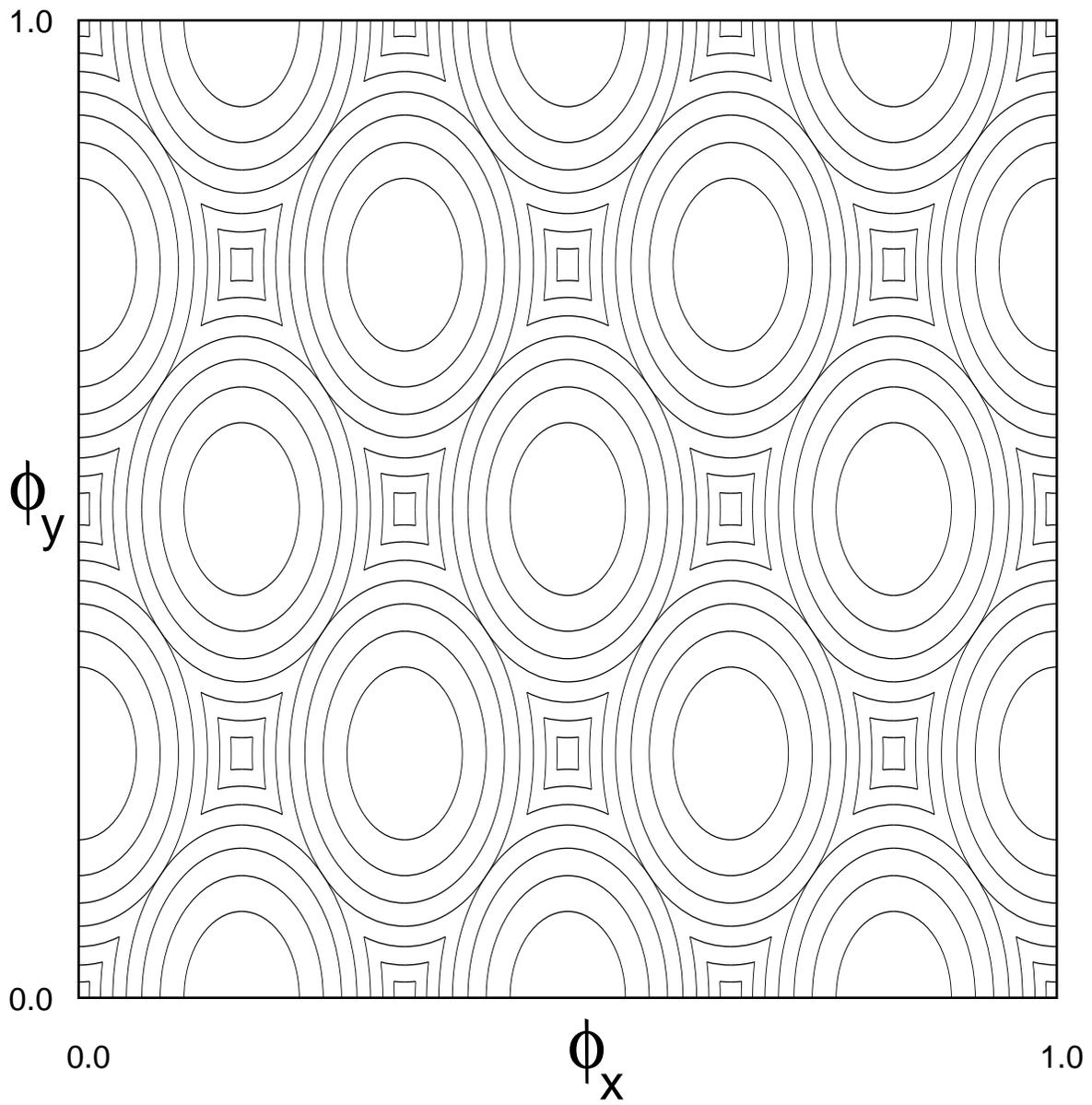
\end{document}